\documentclass[prd,aps,a4paper,superscriptaddress,twocolumn,nofootinbib,floatfix]{revtex4}

\pdfoutput=1
\usepackage{graphicx}
\usepackage{color}
\usepackage{dcolumn}
\usepackage{bm}
\usepackage{slashed}
\usepackage{amsmath}
\usepackage{latexsym}
\usepackage{amssymb}
\usepackage{mathrsfs}
\usepackage{placeins}
\usepackage{booktabs}
\usepackage{subfigure}
\usepackage{float}
\usepackage[normalem]{ulem}
\usepackage{threeparttable}
\usepackage{amsfonts}
\usepackage{wrapfig}
\usepackage{autobreak}
\usepackage{mathtools}
\usepackage{multirow}
\usepackage{array}
\usepackage{graphicx}  
\usepackage{epstopdf}

\allowdisplaybreaks
\usepackage[linktoc=none]{hyperref}
\hypersetup{colorlinks,			
	linkcolor=blue,
	citecolor=blue}
\allowdisplaybreaks
\begin{document}

\title{The effects of data gaps on ringdown signals with space-based joint observation }

\author{Junxi Shi}
\email{shijunxi23@mails.ucas.ac.cn}
\affiliation{International Centre for Theoretical Physics Asia-Pacific, University of Chinese Academy of Sciences, 100190 Beijing, China}

\author{Jiageng Jiao}
\email{jiaojiageng@ucas.ac.cn}
\thanks{Corresponding author}
\affiliation{International Centre for Theoretical Physics Asia-Pacific, University of Chinese Academy of Sciences, 100190 Beijing, China}

\author{Jingqi Lai}
\email{laijingqi19@mails.ucas.ac.cn}
\affiliation{School of Physical Sciences, University of Chinese Academy of Sciences, Beijing 100049, China}

\author{ZhiXiang Li}
\email{lizhixiang21@mails.ucas.ac.cn}
\affiliation{School of Fundamental Physics and Mathematical Sciences, Hangzhou Institute for Advanced Study, UCAS, Hangzhou 310024, China}

\author{Caiying Shao}
\email{shaocaiying@ucas.ac.cn}
\thanks{Corresponding author}
\affiliation{School of Physical Sciences, University of Chinese Academy of Sciences, Beijing 100049, China}

\author{Yu Tian}
\email{ytian@ucas.ac.cn}
\thanks{Corresponding author}
\affiliation{School of Physical Sciences, University of Chinese Academy of Sciences, Beijing 100049, China}

\begin{abstract}
In space-based gravitational wave observatories such as Taiji, LISA, and TianQin, data gaps are inevitable due to mission design, implementation, and the long duration of observations. These data gaps degrade data quality and cause spectral leakage during Fourier transformations. Since ringdown signals are a key scientific objective for these observatories, it is crucial to assess the impact of data gaps on ringdown signal observations. This study employs LISA’s science requirement of maintaining a duty cycle of at least 75$\%$ to evaluate the worst-case impact of data gaps, and uses massive black hole binary catalogs to assess the average effects. Our findings indicate that, on average, data gaps increase parameter estimation errors by approximately 2.1 times for the (2,2) mode and by about 1.6 times for the (3,3) mode. Joint observation is commonly employed to alleviate the impact of data gaps. Similarly, we have evaluated the effects of joint observation with two configurations, Taiji-LISA and Taiji-TianQin, which demonstrate notable mitigation of the effects of data gaps. This work provides a quantitative assessment of data gaps on ringdown signal and highlights the significance of joint observation.

\end{abstract}
\maketitle

\section{Introduction}
The detection of gravitational waves (GWs) by the LIGO-Virgo Collaboration 
\cite{PhysRevLett.116.061102,Abbott_2017,PhysRevLett.116.241103,PhysRevLett.118.221101,PhysRevLett.119.141101,PhysRevX.6.041015}
represents one of the most significant scientific breakthroughs of our time, opening a new window to explore the universe.
Ground-based observatories like LIGO and Virgo are sensitive to frequency band between 10 Hz and several kHz \cite{PhysRevX.9.031040}, while pulsar timing arrays operate in the nHz to $\mu$Hz range \cite{McLaughlin_2013,Kramer_2013,Hobbs_2013,Xu_2023}.
However, the 0.1 mHz to 1 Hz frequency range is currently inaccessible to detectors, and contains a wealth of GW sources including ringdown signal of massive black hole binaries (MBHBs) \cite{Berti:2005ys,Berti:2018vdi,PhysRevD.73.064030}. Ringdown signal is one of the key topics addressed in this paper. To bridge this detection gap, international efforts have led to proposals for space-based GW detection missions, such as Taiji \cite{Hu:2017mde}, LISA \cite{LISA:2017pwj} and TianQin
\cite{TianQin:2015yph}.
 Unlike ground-based detectors, space-based detectors will operate continuously for four years or longer once launched which requires thorough evaluation of potential operational issues.
Additionally, during detector operation, data gaps can arise from various mechanisms and lead to the collection of unusable data. These gaps will compromise detectability and affect data analysis, making it essential to carefully address this issue.

Periodic data gaps are referred to as ``scheduled gaps". Scheduled gaps can result from several factors, such as periodic data transmissions to Earth due to the limited storage capacity of the detector or adjustments in the orientation of antenna caused by changes in the pointing direction of the spacecraft.
Additionally, GW detectors are subject to data interruptions and anomalies during scientific operation as shown by datas from satellites such as LISA Pathfinder \cite{Audley:2019tnx}, Taiji-1 \cite{doi:10.1142/S0217751X21020024}, and TianQin-1 \cite{Luo:2020bls}. 
These interruptions affect both data continuity and quality, resulting in ``unscheduled gaps". While scheduled gaps can be adjusted according to operational requirements, unscheduled gaps cannot yet be systematically modeled. However, unscheduled gaps can be evaluated using a model with a random gap distribution following an exponential distribution based on the observation requirements of space-based GW detectors, ensuring a duty cycle of no less than 75$\%$ \cite{AmaroSeoane2021}.

Previous researchers have constructed data gap models and evaluated the impact on the detection of Galactic Binaries \cite{Carre:2010ra} and inspiral signals from MBHBs \cite{Dey:2021dem}. The results indicate that scheduled gaps have minor effects on the achievement of scientific objectives, whereas the influence of unscheduled gaps is significantly greater. For inspiral signals, unless gaps appear near the merger phase the impact will be negligible \cite{Dey:2021dem}. Dealing with data gaps to relieve the impact is also important. Baghi et al. \cite{baghi3} addressed the issues by using Bayesian data augmentation and Wang et al. \cite{Wang:2024bod} used an inpainting method to cope with data gaps. Their results show significant improvement in parameter estimation (PE) accuracy. 

Ringdown signals from post-merger phase of MBHBs are a significant scientific objective for GWs observation. The measurement of Quasinormal modes (QNMs) not only serves to verify the no-hair theorem \cite{PhysRevD.101.104005}, but also constraints modified theories of gravity \cite{PhysRevD.102.124011,Shao:2023yjx} which are highly valuable. The duration of ringdown signals is very short, lasting from a few seconds to several hours within the sensitive frequency band of the detectors. Given that data gaps occur on a similar timescale to ringdown signals, the impact on observation is expected to be significant. Given the importance of the investigation regarding ringdown signals, it is crucial to thoroughly assess how data gaps affect the observation. The impact of data gaps manifests in the scientific investigation of the corresponding GW sources. This paper presents an analysis of how these data gaps affect the signal-to-noise ratio (SNR) and subsequently the accuracy in testing the no-hair theorem.

Joint observation provides a potential solution to mitigate the impact of data gaps. It allows detectors of different types or operating in different frequency bands to complement each other's strengths \cite{Wang:2021uih,Lyu:2023ctt}. This paper primarily focuses on the joint observation of different space-based GW detectors. Data gaps are independent across detectors. When one detector experiences a data gap, another one may still be operating normally. Thus, joint observation by multiple detectors can mitigate the impact of data gaps on observing the same GW event. This paper considers different configurations to evaluate the advantage of joint observation in addressing data gaps.

This paper is organized as follows: Section \ref{sec:datagap} details data gaps and the construction of signals in the presence of gaps. Section \ref{sec:metho} presents the noise model and ringdown signal model utilized in this study and describes the methods used to assess the impact of data gaps on GW signal analysis. Since joint observation is a key focus, section \ref{sec:joint} provides a brief overview of its role in this study. In section \ref{sec:results}, we present our main findings regarding the detectability of ringdown signals in the presence of gaps and the advantages of joint observation. The analysis is based on Fisher Information Matrix (FIM) and astrophysical catalogs. Finally, section \ref{sec:conclusion} summarizes our conclusions and offers an outlook on future research directions.

\section{ DATA GAPS }
\label{sec:datagap}
The onboard antenna of the detectors is used for communication with Earth. However, the orbital drift of the detector causes changes in the antenna's pointing direction. To ensure normal data transmission, the antenna's orientation must be periodically adjusted. These antenna rotations generate additional noise, making the collected data unusable during this period. This phenomenon is one of the major sources of scheduled data gaps. Additionally, periodic maintenance modes, test mass discharging, and other similar factors also contribute to scheduled data gaps.

The detector is a highly precise and complex machine operating in the harsh space environment. It faces a series of challenges such as high-energy particle impacts on test masses, thermal disturbances, and magnetic interference, all of which can cause unscheduled gaps. During the initial stages of scientific operation, the likelihood of anomalies is relatively low. However, as the mission progresses, the likelihood of anomalies increases, making it reasonable to model the occurrence of data gaps using an exponential distribution. Therefore, the interval $\Delta T$ between two consecutive gaps has a probability density function given by \cite{AmaroSeoane2021}

\begin{equation}
    \frac{dp}{d\Delta T} = \lambda e^{-\lambda \Delta T},
\end{equation}
where $\lambda$ is a parameter of exponential distribution.

By applying a window function to the original time domain signal, we can construct signals including data gaps. Typically, ``applying a window function" refers to a technique for extracting specific information from data. In the context of this paper, this means setting a part of signal to zero over specific time intervals to simulate data gaps. In practice, the window function can be expressed as
\begin{equation}
G(t) =
\begin{cases} 
0,& t^{i}_{gap}<t<t^{e}_{gap}, \\
1,& t<t^{i}_{gap},t>t^{e}_{gap},
\end{cases}
\end{equation}
where $t^{i}_{gap}$ indicates start point and $t^{e}_{gap}$ indicates end point of data gaps, respectively. While the $h(t)$ represents the optimal signal without any data gaps, applying the window function to $h(t)$, then the signal with gaps will be \cite{Carre:2010ra,Dey:2021dem}
\begin{equation}
    h_G (t) = G(t) h(t).
\end{equation}

In the data analysis process, it is often necessary to transform the signal from the time domain to the frequency domain using a Fourier transform. If the aforementioned data gaps are directly applied, the resulting discontinuities will cause significant spectral leakage. To mitigate this issue, $Carr\acute{e}$ et al. \cite{Carre:2010ra} explored the use of different window functions to smooth the data gaps and identify the most suitable one. We adopt a cosine-shaped window function to simulate data gaps, as described in Ref. \cite{Dey:2021dem}:

\begin{equation}
\label{eq:winfunc}
G(t) = 
\begin{cases} 
\frac{1}{2} \left( 1 + \cos \left[ \pi \frac{t - t_s - t_{tr}}{t_{tr}} \right] \right), & t_s - t_{tr} < t < t_s \\
0, & t_s < t < t_e \\
\frac{1}{2} \left( 1 + \cos \left[ \pi \frac{t - t_e - t_{tr}}{t_{tr}} \right] \right), & t_e < t < t_e + t_{tr} \\
1. & \text{otherwise}
\end{cases}
\end{equation}

This window function allows for a smooth transition between regions with and without data gaps, helping mitigate spectral leakage. In this function, $t_{tr}$ represents the transition time, while $t_s$ and $t_e$ denote the start and end times of the data gap, respectively. The transition time can be manually adjusted depending on the analysis requirements. A shorter transition time more accurately reflects real data but may cause some spectral leakage. Conversely, a longer transition time nearly eliminates spectral leakage but causes more signal amplitude to be lost during the transition, thereby lowering the SNR.

Since data analysis in the time domain is computationally intensive and some information is more easily discerned in the frequency domain, this paper conducts the analysis in the frequency domain. To ensure comparability, the signal is first transformed to the time domain via an inverse Fourier transform. After introducing data gaps, the signal is transformed to the frequency domain again via a Fourier transform. Finally, signals with and without data gaps are compared in the frequency domain to evaluate the impact of data gaps
\section{methodology}
\label{sec:metho}
This section introduces the models necessary for the calculations, as well as the metrics used to evaluate the results. We firstly review the noise model for space-based GW observatories and the waveform for ringdown signals. The most direct evaluation metric is the SNR, which measures the strength of the detected signal. A signal with the SNR greater than 8 is generally considered to be successfully detected. The FIM is used for preliminary PE to obtain error bounds; however, more accurate estimates require Bayesian inference. Finally, since results from a single source may not be representative, catalogs based on astrophysical models are used to evaluate the average impact of data gaps over several years of mission duration.
 
\subsection{Noise Model}
\label{sec:noise}
To calculate the SNR and PE errors this paper uses the following noise model and response function \cite{Robson:2018ifk}:

\begin{table*}
\caption{\label{tab:detector_params}Parameters of all three space-based detectors.}
\resizebox{0.9\textwidth}{!}{
\begin{ruledtabular}
\begin{tabular}{lccc}
 & LISA & TianQin & Taiji \\
\hline
$L$ & $2.5 \times 10^9$ m & $\sqrt{3} \times 10^8$ m & $3 \times 10^9$ m \\
$\sqrt{S_a}$ & $3 \times 10^{-15}$ m s$^{-2}$/Hz$^{1/2}$ & $10^{-15}$ m s$^{-2}$/Hz$^{1/2}$ & $3 \times 10^{-15}$ m s$^{-2}$/Hz$^{1/2}$ \\
$\sqrt{S_x}$ & $1.5 \times 10^{-11}$ m/Hz$^{1/2}$ & $10^{-12}$ m/Hz$^{1/2}$ & $8 \times 10^{-12}$ m/Hz$^{1/2}$ \\
\end{tabular}
\end{ruledtabular}
}
\end{table*}

\begin{equation}
\begin{split}
    &S_n(f)= \frac{10}{3L^2} \left( P_{\text{OMS}}(f) + \frac{4 P_{\text{acc}}(f)}{(2 \pi f)^4} \right) \left( 1 + \frac{6}{10} \left( \frac{f}{f_*} \right)^2 \right) 
    \\& \quad + S_c(f), 
    \\&P_{\text{OMS}}(f) = S_x \left[ 1 + \left( \frac{2 \ \text{mHz}}{f} \right)^4 \right] \ \text{Hz}^{-1}, \\
    &P_{\text{acc}}(f) = 
    \\& \quad S_a \left[ \left( 1 + \left( \frac{0.4 \ \text{mHz}}{f} \right)^2 \right) \left( 1 + \left( \frac{f}{8 \ \text{mHz}} \right)^4 \right) \right] \ \text{Hz}^{-1}.
\end{split}
\end{equation}

Here, $L$ represents the arm length of the detector, $f_* = c/2\pi f $ is the transfer frequency, $S_a$ denotes the acceleration noise, and $S_x$ represents the optical metrology noise. The parameters for the LISA, Taiji, and TianQin observatories are listed in Table \ref{tab:detector_params}.

In the Milky Way, there are a large number of unresolved white dwarf binary sources, whose signals contribute to noise around 2 mHz referred to as foreground noise \cite{cornish:confusion}. For LISA, the foreground noise can be fitted using the following formula
\begin{equation}
\label{eq:foreground_noise}
    S_c(f) = Af^{-7/3}e^{-f^a + \beta f \sin(\kappa f)}[1+\tanh(\gamma(f_k - f))] \ \text{Hz}^{-1}.
\end{equation}
Considering a four-year mission duration, the parameter values are as follows: $A = 9\times 10^{-45}$, $\alpha = 0.138$, $\beta = -221$, $\kappa = 521$, $\gamma = 1680$ and $f_k = 0.00113$. The sensitivity curve of Taiji around 2 mHz differs very little from that of LISA, thus the foreground noise is also nearly the same \cite{Liu:2023qap}. Therefore, Taiji similarly adopts the foreground noise fitting form of equation 
(\ref{eq:foreground_noise}). TianQin is more sensitive at higher frequencies, with poorer sensitivity around 1 mHz. As a result, it is less affected by foreground noise, and therefore, this paper does not consider the foreground noise for TianQin. 

\subsection{Ringdown Signal}
\label{sec:rd}
GWs emitted by MBHBs consist of three phases: inspiral, merger, and ringdown. The ringdown signal contributes considerable SNR. The ringdown signals can be modeled as a superposition of several QNMs that oscillate and decay over time. In the framework of general relativity, the QNMs of a Kerr black hole are typically characterized by three indices: $l$, $m$, and $n$, where $n = 0, 1, 2, 3 \ldots$ is the overtone index, and $l$ and $m$ are the harmonic indices. The modes with $n = 0$ are referred to as the fundamental mode, which generally has a larger amplitude and a longer decay time compared to modes with $n \geq 1$, which is easier to be detected. Therefore, in this subsection, we focus only on the fundamental modes, and different QNMs can be characterized simply by $(l, m)$. Although all QNMs contribute to the ringdown, only the dominant modes need to be considered to effectively model the ringdown waveform. We select the $(2,2)$, $(2,1)$, $(3,3)$, and $(4,4)$ modes to construct the ringdown signals.

Ringdown waveform can be represented as 
\begin{equation}
\begin{split}
\label{eq:ringdown_waveform}
    h_{+ , \times}(t) &= \frac{M_z}{D_L} \sum_{l,m>0} A_{lm} Y^{lm}_{+ , \times}(\iota) \Psi^{+ , \times}_{lm}(t),\\
    \Psi^{+}_{lm}(t) &= \exp\left(-\frac{t}{\tau _{lm}}\right) \cos\left(\omega_{lm} t - m \phi_{lm}\right), \\
    \Psi^{\times}_{lm}(t) &= -\exp\left(-\frac{t}{\tau_{lm}}\right) \sin\left(\omega_{lm} t - m \phi _{lm}\right)
\end{split}
\end{equation}
for $t \geq t_0$, $h_{+,\times}(t) = 0$ for $t<0$. Where $t_0$ is the starting point of the ringdown phase, $M_z$ is the redshifted mass of the remnant black hole, and $A_{lm}$, $\omega_{lm}$, $\tau_{lm}$, and $\phi_{lm}$ represent the amplitude, oscillation frequency, damping time, and initial phase of the QNMs, respectively. Fitting formulas for $A_{lm}$ have been provided by \cite{Baibhav:2017jhs,Baibhav:2018rfk}. $\iota$ is the inclination angle of the source, where $\iota \in [0, \pi]$. The function $Y_{lm}(\iota)$ can be expressed as a linear combination of spin-weighted spherical harmonics with spin weight $-2$ \cite{Kamaretsos:2011um}:
\begin{equation}
    \begin{split}
        Y^{lm}_+(\iota) = \prescript{}{2}{Y}^{lm}(\iota,0) + (-1)^{l} \prescript{}{-2}{Y}^{l-m}(\iota,0),\\
        Y^{lm}_\times (\iota) = \prescript{}{2}{Y}^{lm}(\iota,0) - (-1)^{l}\prescript{}{-2}{Y}^{l-m}(\iota,0).\\
    \end{split}
\end{equation}
$\omega_{lm}$ and $\tau_{lm}$ in equation (\ref{eq:ringdown_waveform}) can be given by the following equation \cite{Berti:2016lat}:
\begin{equation}
\label{eq:freq}
    \begin{split}
        \omega_{lm} &= \frac{f_1 + f_2 (1-\chi_f)^{f_3}}{M_z},\\
        \tau_{lm} &= \frac{2(q_1 + q_2 (1-\chi_f)^{q_3})}{\omega_{lm}}.
    \end{split}
\end{equation}
Fitting coefficients of this equation are provided in Table \ref{table:coefficients}, and $\chi_f$ is angular momentum of remnant black hole. For the merger of black hole without spin, $\chi_f$ is only dependent by mass ratio which can be expressed by \cite{Barausse_2009}
\begin{equation}
    \chi_f (q) = \eta(2\sqrt{3}-3.517 \eta + 2.5763 \eta^2),
    \label{eq:spin}
\end{equation}
where $\eta = q/(1+q)^2$ is  symmetric mass ratio. 

\begin{table}[htbp]
\centering
\caption{\label{table:coefficients}The fitting coefficients for equation (\ref{eq:freq}). Taken from Ref. \cite{Berti:2005ys}.}
\begin{tabular}{cccccccc}
\hline
\hline
$(\ell, m)$ & $f_1$  & $f_2$  & $f_3$  & $q_1$  & $q_2$  & $q_3$  \\
\hline
(2, 2) & 1.5251  & -1.1568  & 0.1292  & 0.7000  & 1.4187  & -0.4990  \\
(3, 3) & 1.8956  & -1.3043  & 0.1818  & 0.9000  & 2.3430  & -0.4810  \\
(2, 1) & 0.6000  & -0.2339  & 0.4175  & -0.3000 & 2.3561  & -0.2277  \\
(4, 4) & 2.3000  & -1.5056  & 0.2244  & 1.1929  & 3.1191  & -0.4825  \\
\hline
\hline
\end{tabular}
\end{table}

The detected signals depend not only on the parameters of the ringdown phase but also on the orientation of the sources relative to the detector and the properties of the detector itself. A space-based GW observatory consists of three spacecraft arranged in an equilateral triangle configuration, with each pair of spacecraft connected by two laser arms. This type of detector can be considered as equivalent to two LIGO-like detectors (denoted as ``I" and ``II") with an opening angle of $\gamma = \pi/3$.

In the detector's coordinate system, the orientation and polarization angle of the source are represented by $(\theta_d, \phi_d)$ and $\psi_d$, respectively. The signal from a source located in the direction $(\theta_d, \phi_d)$ in detector I is equivalent to the signal from the same source located in the direction $(\theta_d, \phi_d + 2\pi/3)$ for detector II. For details on calculating $\theta_d$, $\phi_d$, and $\psi_d$, as well as the specific form of the parameters in equation (\ref{eq:sfa}) related to the detected signal, please refer to Ref. \cite{Zhang:2021kkh}.

Doppler frequency shifts in the detected signals are scalar quantities. Therefore, in theoretical analysis, the waveform of different polarization modes of the GW is contracted with the detector tensor to obtain a scalar signal. Furthermore, to save computational resources, the construction and calculation of the signal will be carried out in the frequency domain. The corresponding signal can be written as
\begin{equation}
\label{eq:sfa}
    s(f) = \sum_{A = +,\times}[D^{A}_{u} \mathcal{T}(f,\hat{u} \cdot \hat{o} ) - D^{A}_{v}\mathcal{T}(f,\hat{v} \cdot \hat{o})]h_{A}(f),
\end{equation}
where $D$ represents the detector tensor, $\mathcal{T}$ is the transfer function, $A$ denotes the GW polarization mode, and $h_A(f)$ is the signal in the frequency domain for a particular polarization mode. Ringdown signals have a very short duration. For mHz-range signals, the shortest duration is only a few seconds, while the longest does not exceed one day. The rotational frequency of heliocentric detectors is extremely low; for example, the rotational frequency of LISA is $\omega_{\text{LISA}} = 1.99 \times 10^{-7}$. Therefore, the rotation of heliocentric detectors like LISA can be neglected. Equation (\ref{eq:sfa}) neglects the effects of rotation. However, for geocentric detectors like TianQin, the higher rotational frequency requires considering the detector's rotation when calculating signals from more massive sources with longer durations.

\subsection{Signal to Noise Ratio}
The SNR is a metric of the strength of the GW signals in the data. Define an inner product operation as \cite{Robson:2018ifk,Finn,Moore:2014lga}
\begin{equation}
    \left(A(f)\middle|B(f)\right) = 4\mathcal{R}e\left(\int^{\infty}_{0}  \frac{A^*(f) B(f)}{S_n(f)} df \right),
\end{equation}
then, SNR $\rho$ can be written as 
\begin{equation}
    \rho = \sqrt{(h(f)|h(f))},
\end{equation}
where $h(f)$ denotes GW signals in the frequency domain. If calculating the SNR in the presence of data gaps, $h(f)$ should be replaced with $h_G(f)$. A detector with an equilateral triangle configuration can be considered equivalent to two L-shaped detectors, allowing the total SNR to be expressed as
\begin{equation}
    \rho^2_{total} = \rho^2_1 + \rho^2_2.
\end{equation}

\subsection{Fisher Information Matrix}
Data loss inevitably leads to increased PE errors. The accurate way to determine the magnitude of the errors is to use Bayesian inference to calculate the posterior distribution for each GW source. However, Bayesian inference is computationally expensive, making it impractical to apply to all sources in the simulated catalogs. An alternative approach is to calculate the FIM. The inverse of the FIM gives the Cramer-Rao bound \cite{valli}, which provides a lower bound for the covariance of the source parameters in the high SNR regime. However, the Cramer-Rao bound is only achievable in the high SNR regime.

Therefore, for low SNR signals from low-mass systems, the error estimates from the FIM tend to be optimistic \cite{Nicho}. Despite this limitation, the FIM provides a useful baseline for further research, allowing us to manage computational costs effectively before analyzing a large number of sources. It also helps establish a limit on the best achievable accuracy of PE.

In the high SNR limit, the stand deviation of PE can be calculated by
\begin{equation}
\label{eq:inverseFIM}
    \Delta \theta^{\alpha} = \sqrt{\langle \delta \theta^{\alpha} \delta \theta^{\alpha} \rangle} = \sqrt{(\Gamma^{-1})^{\alpha \alpha}},
\end{equation}
where $\theta^\alpha $ represents a parameter to be evaluated, and $\langle \ldots \rangle$ denotes the expectation value. $\Gamma^{-1}$ is the covariance matrix, which is also the inverse of the FIM. The FIM is defined as
\begin{equation}
\label{eq:fisher_matrix}
    \Gamma_{ab} = \left(\frac{\partial h}{\partial \theta^{a}}\middle| \frac{\partial h}{\partial \theta^b}\right).
\end{equation}
Similarly, $h$ should be replaced with $h_G$ when calculating signals with data gaps.

The main issue considered in this paper is testing the no-hair theorem through ringdown, specifically, analyzing potential deviations of the QNMs from the values predicted by general relativity. Therefore, to perform PE, the frequencies of the QNMs in equation (\ref{eq:sfa}) must be replaced by \cite{gossen,Li2012}
\begin{equation}
    \begin{split}
        \omega_{lm} &= \omega_{lm,\text{GR}}(1+\delta \omega_{lm}),\\
        \tau_{lm} &= \tau_{lm,\text{GR}}(1+\delta \tau_{lm}).
    \end{split}
\end{equation}
If the estimated values of all $\delta \omega_{lm}$ and $\delta \tau_{lm}$ are zero, it indicates that the predictions of general relativity are correct. Contrarily, If the frequency of any mode shows sufficient deviation, then general relativity may need correction. We focus on the deviation of QNMs and the parameter space to be considered is
\begin{equation}
    \vec{\theta} = \{M_z,  \eta, D_L, \iota, \theta_d, \phi_d, \psi_d, t_0, \phi_{lm}, \delta \omega_{lm}, \delta \tau_{lm} \}.
\end{equation}

\subsection{Astrophysical Catalogs}

\begin{figure}[htbp]
    \centering
    \includegraphics[width=0.45\textwidth]{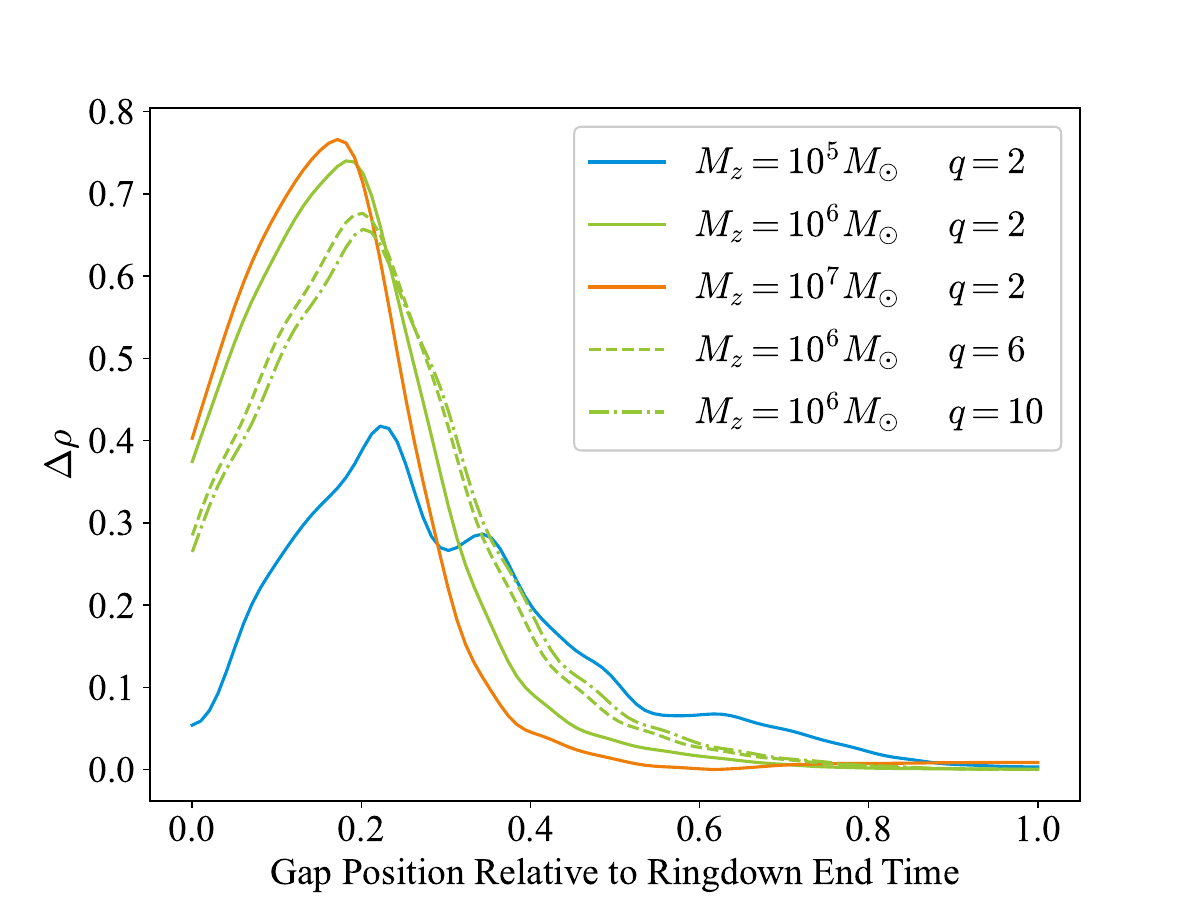}
    \caption{Plot of $\Delta \rho$ as a function of the position of a single data gap. To illustrate the impact of data gaps, a single gap with a 75$\%$ duty cycle was used. The horizontal axis represents the position of the gap center, where ``0" corresponds to the start of the ringdown and ``1" to the end. The vertical axis displays the relative change $\Delta \rho$ caused by the data gap. The position of the data gap significantly affects the $\rho$, with larger remnant mass sources being more impacted by the gap compared to smaller mass sources.}
    \label{fig:snr_gap_position}
\end{figure}

As shown in Fig. \ref{fig:snr_gap_position}, the impact of data gaps varies significantly with their position. Therefore, the result for a single event is subject to significant randomness, making the average results obtained from astrophysical models more reliable. To calculate the average results, we focus on two main classes of seed models: the PopIII light seed model and the Q3-delays and Q3-nodelays heavy seed models, each of which represents a different mechanism for black hole formation in the early universe and brings different expected results.

The PopIII model posits that massive black hole (MBH) seeds formed from the remnants of Population III stars, which are believed to have formed in low-metallicity environments at redshifts $z\sim15-20$. These early stars were likely massive, but those within the mass range of $140 - 260 M_\odot$ would have undergone pair-instability supernovae, leaving no black hole remnants. However, stars outside this range could leave behind black holes with masses around two-thirds of the initial stellar mass, typically resulting in light seed black holes of $100 - 300 M_{\odot}$. This model predicts lighter MBH seeds, which evolve through accretion and mergers to form the MBHs observed in the local universe. Consequently, the model forecasts relatively low-mass binary black hole systems, which would generate weaker gravitational wave signals with lower signal-to-noise ratios, making detection more challenging. The PopIII model provides a framework for exploring the growth of MBHs from smaller seeds, although it predicts less optimistic event rates compared to the heavy seed models \cite{klein2017, Madau_2001, Heger_2002}.

The Q3-delays and Q3-nodelays models represent heavy seed formation scenarios \cite{klein2017, barausse2, Antonini_2015, hoffman}. These models assume that MBH seeds formed at high redshifts $z \sim 15 - 20$ through dynamical instabilities in protogalactic disks. In this scenario, large amounts of cold gas are funneled into the nuclear regions of galaxies, where the gas collapses to form black hole seeds with masses on the order of $10^{4} - 10^{5} M_{\odot}$. These heavier seeds can rapidly grow into supermassive black holes through accretion and mergers.

The Q3-delays model accounts for delays between galaxy mergers and the subsequent MBH mergers. After galaxies merge, the black holes they host must migrate to the center of the newly formed galaxy via dynamical friction, a process that can take billions of years. Consequently, MBH mergers in the Q3-delays model are expected to occur later in the evolution of universe, typically at lower redshifts. This model provides a more conservative prediction of MBHB formation, reflecting a realistic scenario in which black holes take time to coalesce after galaxy mergers.

In contrast, the Q3-nodelays model assumes that MBH mergers occur without significant delays following galaxy mergers. In this scenario, black holes in merging galaxies can quickly form binary systems and coalesce, avoiding the prolonged timescales associated with dynamical friction and other processes. As a result, this model predicts more frequent MBH mergers at higher redshifts, leading to a higher event rate of MBHBs.

\section{Joint observation}
\label{sec:joint}
Joint observation is a key focus of this paper as it helps mitigate the effects of data gaps. Data gaps arise from operational schedules and unavoidable detector malfunctions so that they occur independently in different detectors. Therefore, one detector's data gap may occur at a position that severely affects the detectability, while another detector's data gap may appear in a position that has little to no impact on the detectability. In such cases, two independent detectors can cross-verify the source parameters through their respective signals, thereby reducing the impact of data gaps.

Of course, once the detectors are operational, the ideal solution to address data gaps is to reconstruct the lost data. This not only improves the SNR but also avoids complications like spectral leakage during data processing. This is the subject of ongoing research by our team. However, this does not diminish the importance of joint observation, because there are inevitable errors when reconstructing lost data. If joint observation can provide two or more sets of data, the reconstructed data will be more accurate. Moreover, the process of data reconstruction relies on existing theoretical models, and without the ability to cross-verify signals from multiple detectors, the reconstructed signal might miss certain information that goes beyond the scope of present theoretical models.

The advantages of joint observation for PE have been studied before. Omiya and Seto\cite{omiya_seto}, Seto \cite{seto2020}, and Orlando et al.\cite{Orlando_2021} evaluated network capabilities for the Stochastic GW Background observation. Wang et al.\cite{gangwang2021} estimated the impact of the joint LISA-Taiji observation on PE. Their primary interest lies in understanding effects of joint observation on the complete signals and we will investigate the extent to which it can mitigate the impact of data gaps.

The SNR of a source detected by multiple GW detectors or a GW network can be thus estimated as\cite{Zhao:2023ilw}
\begin{equation}
    \rho^2 = \sum^{n}_{j=1} \mathcal{R}e\left(\int^{\infty}_{0} \frac{4h^*_j(f) h_j(f)} {S_{n,j}}df\right),
\end{equation}
where $j$ represents independent detectors and $n$ refers to the total number of Michelson interferometers in the detector network. Similarly, $h(f)$ should be replaced by $h_G(f)$ calculating the SNR with data gaps. The calculation of PE errors in joint observation is done by first computing the FIM for each individual detector, and then linearly summing the FIM from different detectors as\cite{Zhao:2023ilw}
\begin{equation}
    \Gamma_{ab} = \sum^n_{j=1}\Gamma_{ab,j}.
\end{equation}
Finally, the calculation is completed using equation (\ref{eq:inverseFIM}).

\section{Results}
\label{sec:results}
This section demonstrates the impact of data gaps on ringdown signal observations by assessing the reduction in accuracy of the no-hair theorem test, along with preliminary results on the effectiveness of joint observation in mitigating these impacts. The characteristics of data gap and joint observation are presented in subsection \ref{sec:results1}, while a more comprehensive analysis using astrophysical models is discussed in subsection \ref{sec:resultB}.
\begin{figure}[htbp]
    \centering
    \includegraphics[width=0.45\textwidth]{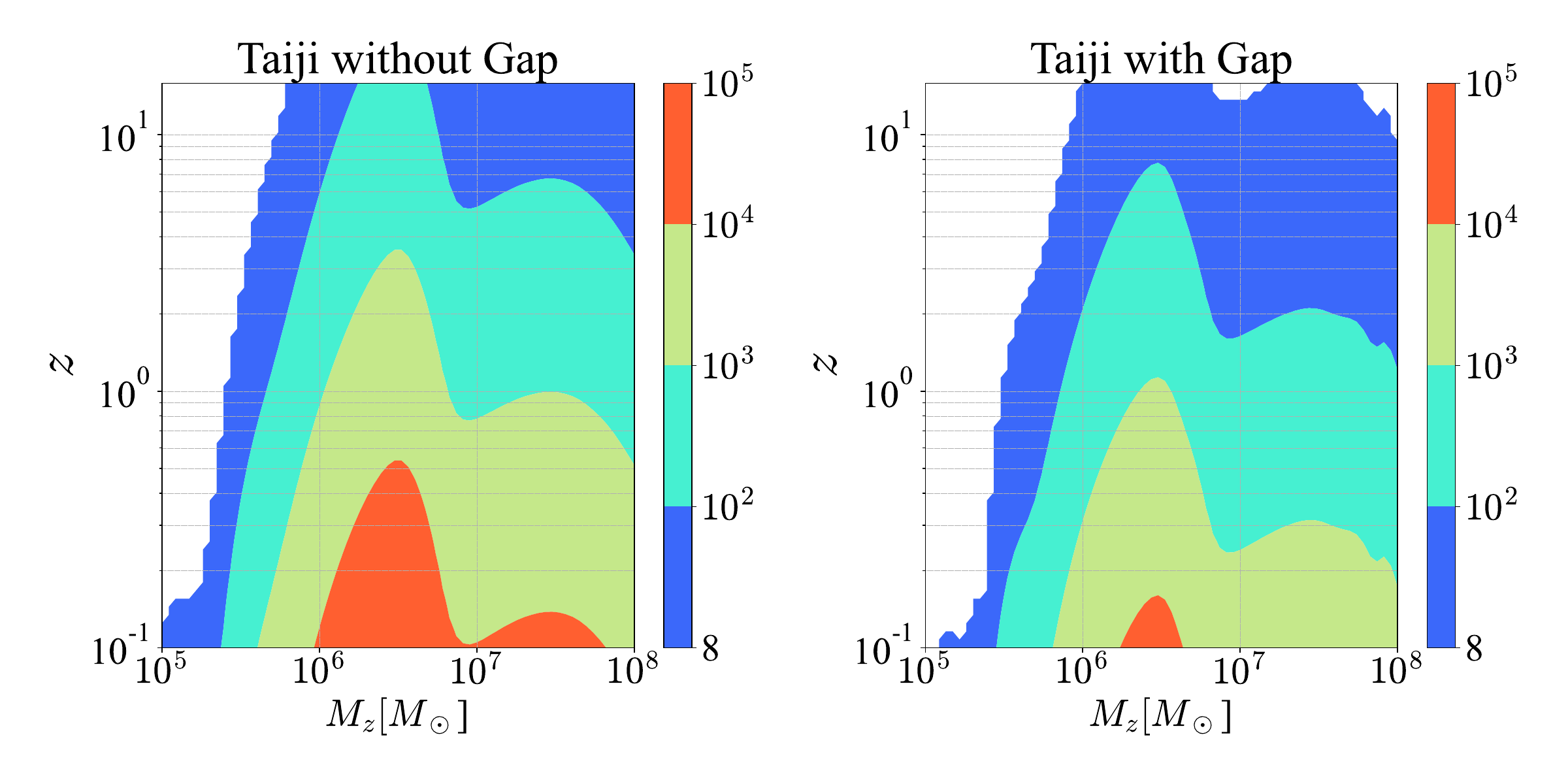}
    \caption{The effects of data gaps on the detection range of ringdown signals in the worst-case scenario for events with a fixed parameter $\eta = 2/9$. The contour indicates corresponding SNR range. Using the Taiji observatory as an example, the left panel illustrates the scenario without data gaps, while the right panel shows the scenario with data gaps.}
    \label{fig:tjmass_range}
\end{figure}

\begin{figure}[htbp]
    \centering
    \includegraphics[width=0.45\textwidth]{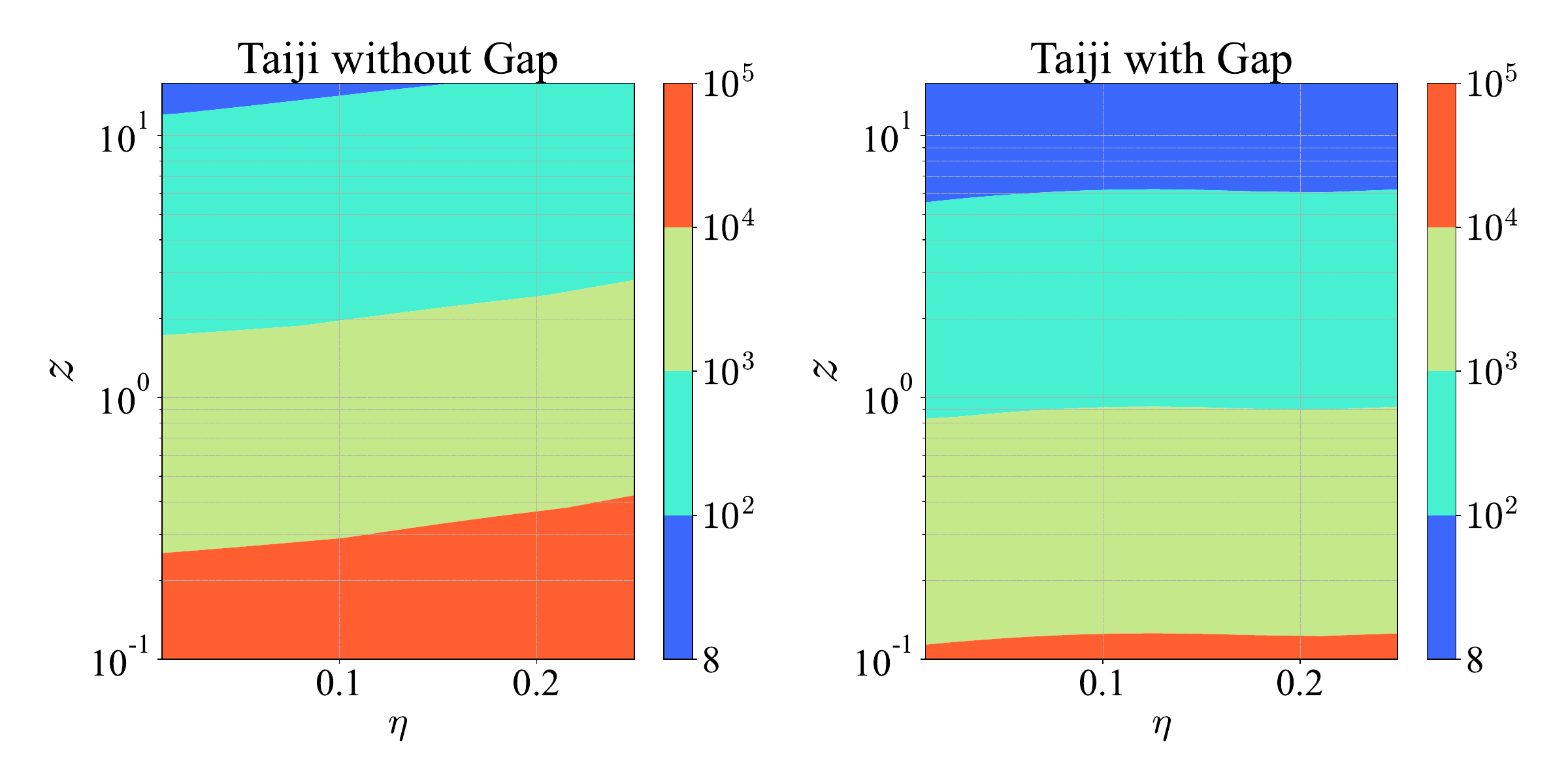}
    \caption{Similar with Fig. \ref{fig:tjmass_range} but for fixed parameter $M_z = 2 \times 10^6 M_\odot$.}
    \label{fig:tjratio_range}
\end{figure}

\begin{figure*}[htbp]
    \centering
    \includegraphics[width=1.0\textwidth]{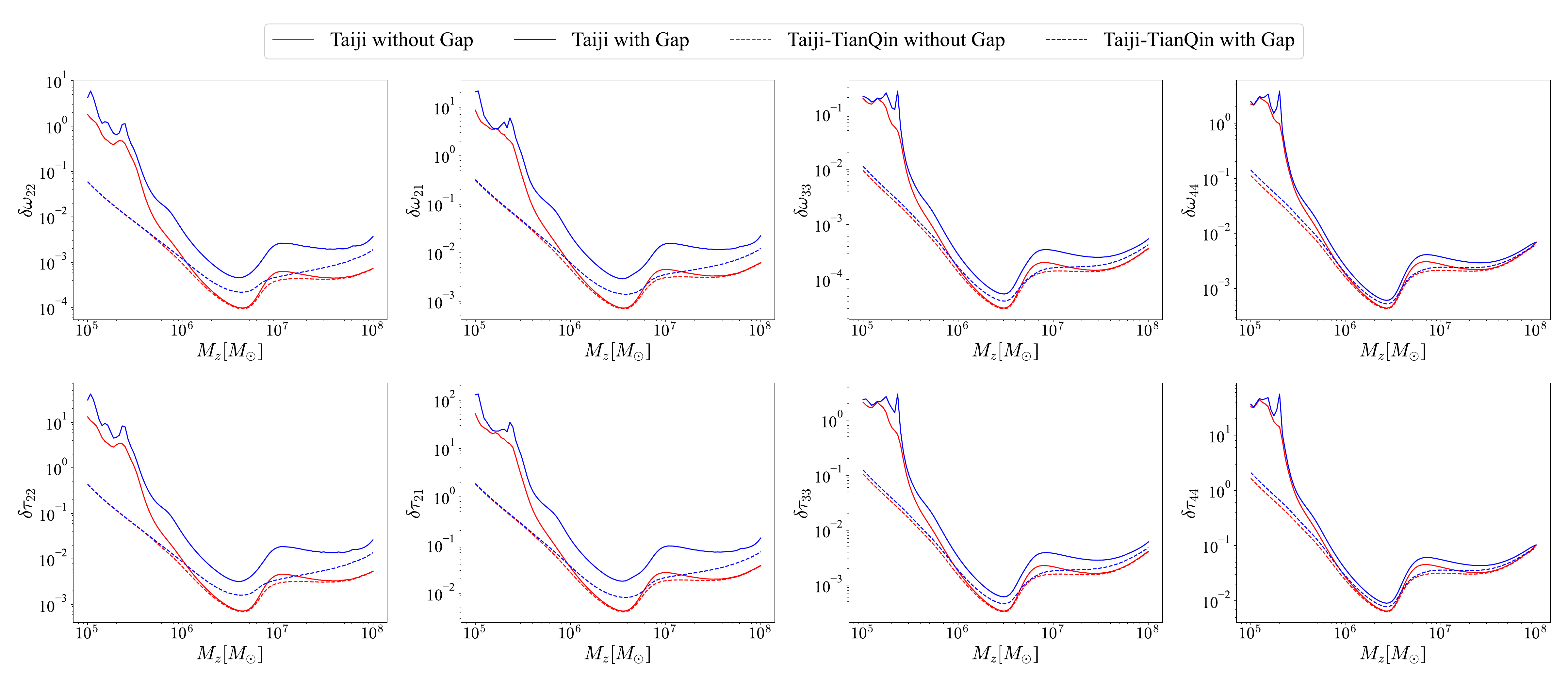}
    \caption{The PE errors for each QNM is evaluated as a function of the remnant mass, with other parameters fixed at $\eta = 2/9$ and $z = 1$. The red and blue solid lines represent the PE errors for a single Taiji observatory with and without the influence of data gaps, respectively. The red and blue dashed lines represent the PE errors for joint observation with Taiji and TianQin. The joint observation assumes that the data gaps of Taiji is in a position with significant impact, while that of TianQin is in a position with minimal impact.}
    \label{fig:tj_deltamass}
\end{figure*}

\begin{figure*}[htbp]
    \centering
    \includegraphics[width=1.0\textwidth]{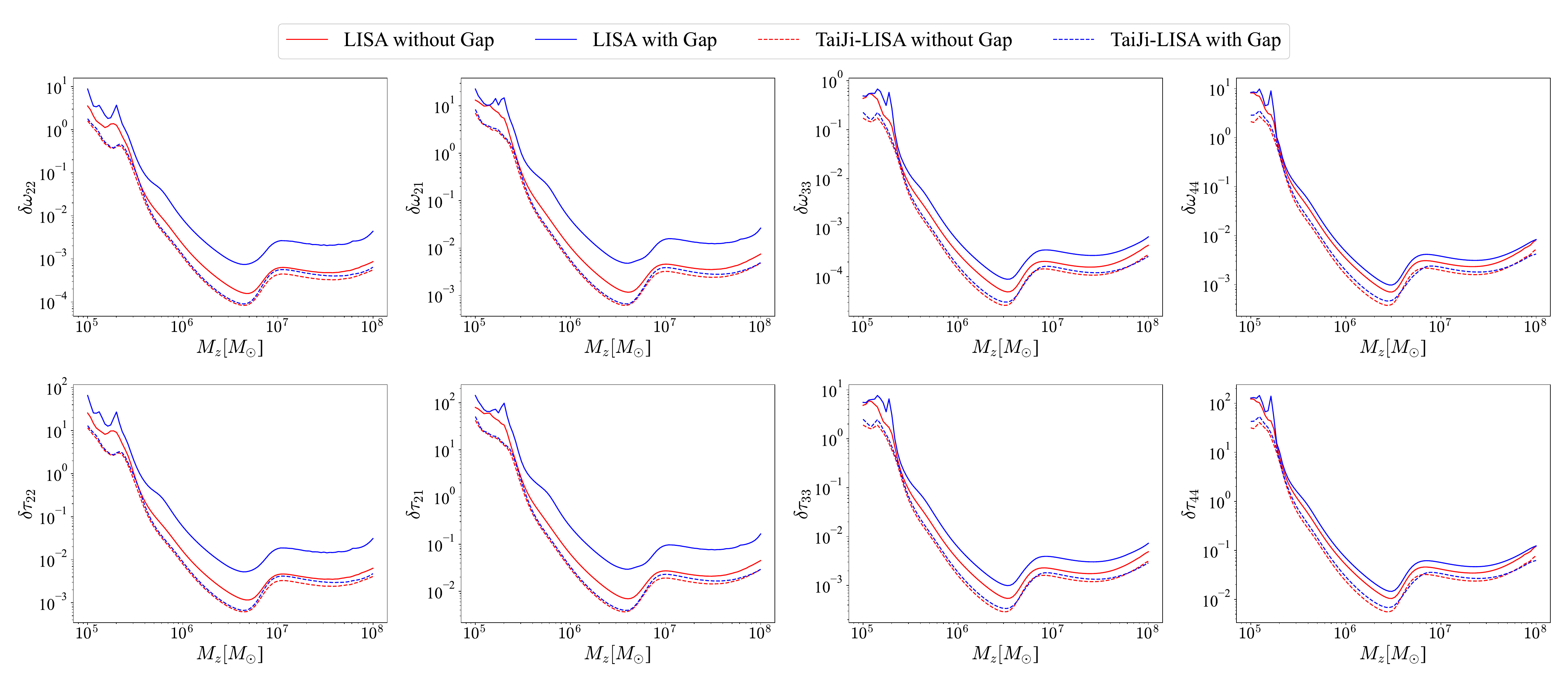}
    \caption{Similar with Fig. \ref{fig:tj_deltamass} but for Taiji and LISA.}
    \label{fig:tj_deltaratio}
\end{figure*}

\subsection{Characteristic Effects of Data Gaps and Joint Observation}
\label{sec:results1}

For simplicity, the orientation parameters of the source in the solar coordinate system are set to $\theta_s = \phi_s = \psi_s = \pi/3$, with the inclination angle set to $\iota = \pi/4$. The initial phase $\phi_{lm}$ of all QNMs and $t_0$ are both set to 0. The severity of the impact of data gaps on a single signal is significantly influenced by randomness. Fig.\kern+0.3em \ref{fig:snr_gap_position} illustrates the effects of a single data gap's position on the detected SNR, using a heliocentric orbit detector as an example. The source mass and mass ratio are varied, while all other parameters remain constant. A single gap with a 75$\%$ duty cycle is injected, with the variable being the position of the gap's midpoint relative to the start of the ringdown signal. On the horizontal axis, “0” indicates that the midpoint of the gap is precisely at the start of the ringdown, while “1” indicates it is near the end of the ringdown. The vertical axis represents the relative change in the corresponding SNR, defined as $\Delta \rho = (\rho_{\text{opt}} - \rho_{\text{gap}})/\rho_{\text{opt}}$, where $\rho_{\text{opt}}$ is the optimal SNR without any gaps and $\rho_{\text{gap}}$ represents the SNR with gaps. The index “opt” is used in this paper to indicate the case without any data gaps.

\begin{table*}
\caption{\label{tab:comparison}Average $r(\theta)$ for different astrophysical models (PopIII, Q3-delays, Q3-nodelays) under various observation configurations (Taiji, Taiji-TianQin, LISA, Taiji-LISA).}
\renewcommand{\arraystretch}{1.5}
\resizebox{0.9\textwidth}{!}{
\begin{ruledtabular}
\begin{tabular}{cccccc}
\textbf{$r(\theta)$} & \textbf{Model} & \textbf{Taiji} & \textbf{Taiji-TianQin} & \textbf{LISA} & \textbf{Taiji-LISA} \\
\hline
\multirow{3}{*}{Average $r(\omega_{ 22}) $} & PopIII & 2.153 & 1.549 & 2.093 & 1.429 \\
 & Q3-delays & 2.162 & 1.521 & 2.136 & 1.468 \\
 & Q3-nodelays & 2.101 & 1.495 & 2.086 & 1.430 \\
\hline
\multirow{3}{*}{Average $r(\tau_{22})$} & PopIII & 2.152 & 1.549 & 2.093 & 1.432 \\
 & Q3-delays & 2.168 & 1.520 & 2.134 & 1.465 \\
 & Q3-nodelays & 2.100 & 1.486 & 2.083 & 1.429 \\
\hline
\multirow{3}{*}{Average $r(\omega_{33})$} & PopIII & 1.608 & 1.359 & 1.601 & 1.296 \\
 & Q3-delays & 1.662 & 1.377 & 1.646 & 1.346 \\
 & Q3-nodelays & 1.622 & 1.350 & 1.629 & 1.322 \\
\hline
\multirow{3}{*}{Average $r(\tau_{33})$} & PopIII & 1.606 & 1.358 & 1.600 & 1.296 \\
 & Q3-delays & 1.661 & 1.377 & 1.645 & 1.345 \\
 & Q3-nodelays & 1.624 & 1.350 & 1.630 & 1.323 \\
\end{tabular}
\end{ruledtabular}
}
\end{table*}

\begin{table}[htbp]
\centering
\caption{\label{table:detection_rate} The proportion of detectable sources that become undetectable due to the impact of data gaps.}
\begin{tabular}{>{\centering\arraybackslash}p{2.4cm} >{\centering\arraybackslash}p{1.8cm} >
{\centering\arraybackslash}p{1.8cm} >
{\centering\arraybackslash}p{2cm}}

\hline
\hline
Models & PopIII  & Q3-delays & Q3-nodelays \\
\hline
Taiji &11.195$\%$&7.486$\%$ &12.459$\%$\\
LISA &12.914$\%$&9.496$\%$&14.356$\%$\\
TainQin &13.111$\%$&5.767$\%$&6.857$\%$\\
Taiji-LISA &8.707$\%$&5.766$\%$&8.696$\%$\\
Taiji-TianQin &10.098$\%$&6.134$\%$   &3.605$\%$\\

\hline
\hline
\end{tabular}
\end{table}

From 
Fig.\kern+0.3em \ref{fig:snr_gap_position}, it is evident that the impact of data gaps depends significantly on their position relative to the signal. A data gap at the start of the ringdown has a notable impact on SNR whereas a gap near the end has little impact. Larger remnant masses are more affected by data gaps. For instance, sources with $M_z \sim 10^7 M_\odot$ experience a $\Delta \rho$ of nearly 80$\%$ at maximum, whereas sources with $M_z \sim 10^5 M_\odot$ show a maximum of around 40$\%$. The effects also vary slightly with mass ratio, being greater for smaller mass ratios.  To more clearly illustrate the impact of data gaps in following results, the gaps in subsection \ref{sec:results1} are placed at the position where the effects are the most significant.

Fig. \ref{fig:tjmass_range} illustrates the impact of data gaps on the detection range of sources with different masses when the symmetric mass ratio $\eta$ is fixed at 2/9. In each panel, the horizontal axis denotes the remnant black hole mass, while the vertical axis represents the redshift. The color of the contour map indicates the corresponding SNR. By comparing the two logarithmic plots on the left and right, we observe that the detection range is significantly reduced by data gaps. The reduction in the contour plot around $10^7 M_\odot$ is due to foreground noise. In Fig. \ref{fig:tjratio_range}, the mass is set to $2 \times 10^6 M_\odot$ to illustrate the impact of data gaps on sources with varying angular momenta. Correspondingly, the horizontal axis in the subplots has been adjusted to represent the symmetric mass ratio. The left panel demonstrates that as the symmetric mass ratioincreases, the strength of the ringdown signal increases, thereby extending the detection range. However, in the right panel, it becomes apparent that the enhancement in detection range within $\eta \sim 0.15 - 0.25$ is flattened due to the impact of data gaps. Thus, data gaps have a more pronounced impact on ringdown signals with higher symmetric mass ratios.

The PE errors of QNM frequencies are crucial for testing the no-hair theorem. Figures \ref{fig:tj_deltamass} and \ref{fig:tj_deltaratio} show the variation in PE errors with remnant mass, fixing the parameters symmetric mass ratio $\eta \sim 2/9$, and redshift $z \sim 1$. Variations with the symmetric mass ratio or remnant spin are not presented, as they are negligible,  consistent with the context and previous studies \cite{Zhang:2021kkh,Shi:2019hqa}. In this case, the impact of data gaps still strongly depends on remnant mass rather than remnant spin.

To verify the effectiveness of joint observation in mitigating the impact of data gaps, Figures \ref{fig:tj_deltamass} and \ref{fig:tj_deltaratio} compare PE errors across different joint observation configurations. In each panel, red and blue lines indicate the absence or presence of data gaps. Solid blue lines represent the results for a single detector affected by data gaps under the worst-case scenario. Dashed lines show the results for joint observation, assuming that the data gap in Taiji is positioned at a highly impactful point, while those in TianQin or LISA are positioned at points with minimal impact.

Comparing the results for the single Taiji or TianQin detector in each panel reveals that in the worst-case scenario, data gaps can affect the (2,2) and (2,1) modes of ringdown signals with masses greater than $10^6 M_\odot$ by nearly an order of magnitude, while the impact on the (3,3) and (4,4) modes is significantly smaller. Comparison of each dushed line shows that due to TianQin’s superior sensitivity to the higher frequency band (see Fig. \ref{fig:tj_deltamass}), joint observation with Taiji notably reduces PE errors for lower mass signals, making the differences almost negligible. More encouragingly, in the case of joint observation by Taiji and LISA (see Fig. \ref{fig:tj_deltaratio}), the PE errors with data gaps show only minor differences compared to the optimal case across the entire range of remnant masses.

These results indicate that joint observation significantly reduces the impact of data gaps on the accuracy of ringdown PE. Although this study focuses on ringdown signals, the methods used are general, and the characteristics of data gap effects align with previous studies. Thus, the conclusions regarding joint observation are also applicable to other types of GW sources. Joint observation can be considered an effective approach to mitigate the effects of data gaps, which will be further discussed in section \ref{sec:resultB}.

\subsection{Average Effects}
\label{sec:resultB}

To provide a more practical assessment of the impact of data gaps, this section evaluates the average effects of data gaps using catalogs from three astrophysical models. Our calculations traverse all sources with an SNR greater than 8 in the catalog of each model. Detection rates and PE errors are computed for scenarios both with and without random data gaps, under both single detector and joint observation configurations. This paper adopts $r(\theta) = \delta \theta_{gap} / \delta \theta_{opt}$ to indicate the impact of gaps. When $r(\theta) = 1$, it indicates no effects from data gaps on PE errors; the larger the value of $r(\theta)$, the greater the impact.

\begin{figure*}[htbp]
    \centering
    \includegraphics[width=0.95\textwidth]{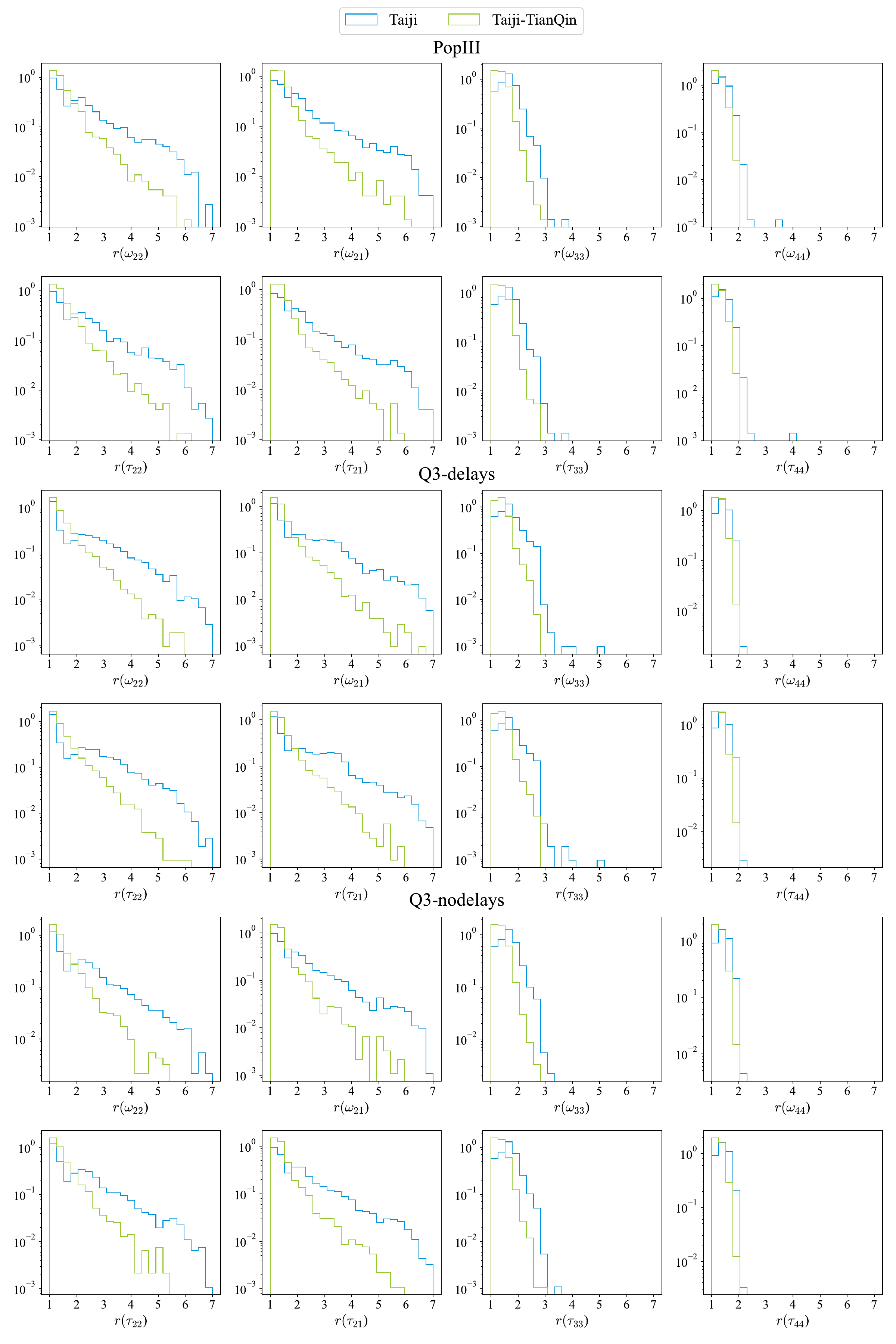}
    \caption{Probability distribution of the ratio of errors $r (\theta)$ between gapped and optimal cases for the sources in three astrophysical models. Blue and green contour indicate the single Taiji detector and the joint Taiji-TianQin configuration respectively.}
    \label{fig:tjtq_catalog}
\end{figure*}

\begin{figure*}[htbp]
    \centering
    \includegraphics[width=0.95\textwidth]{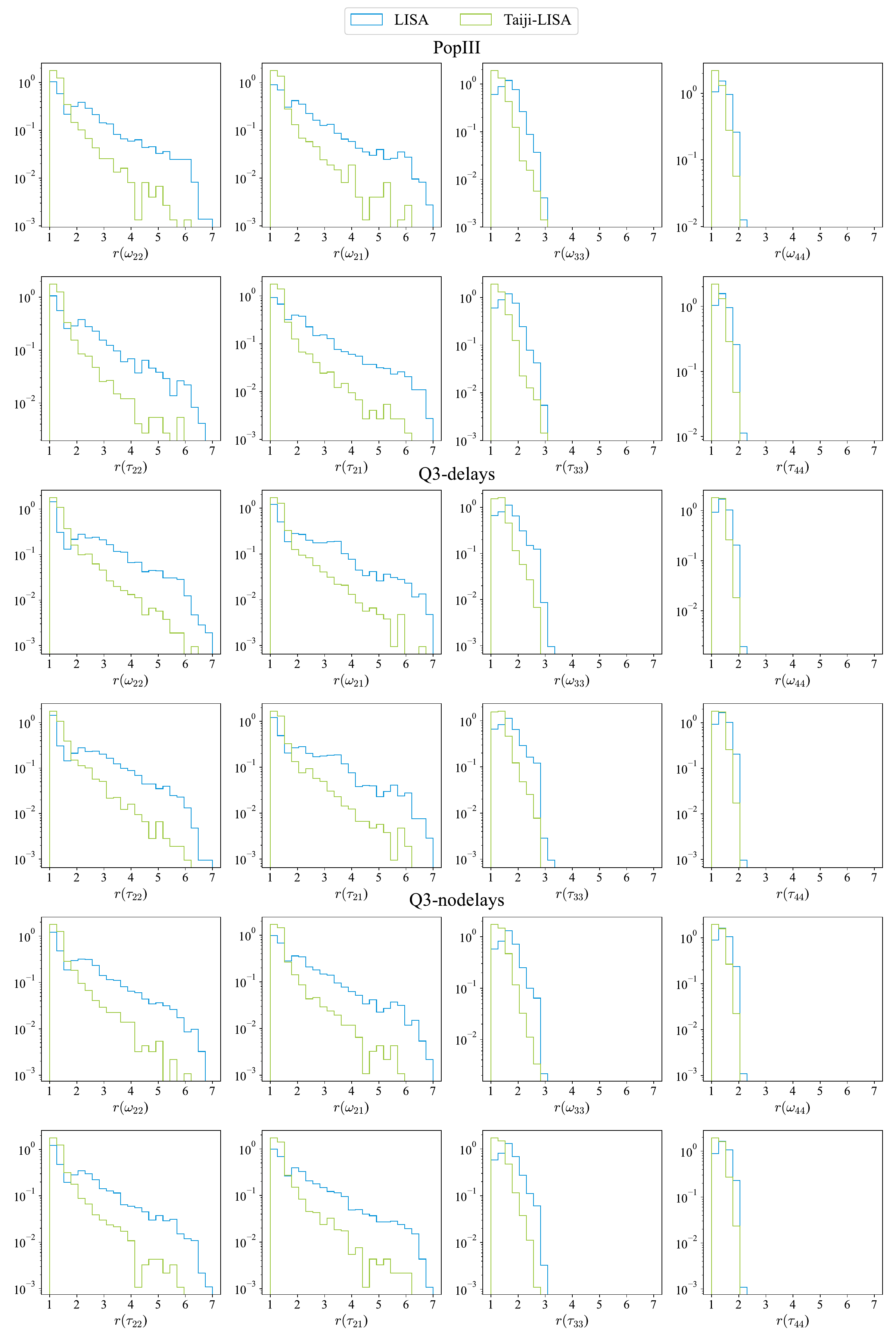}
    \caption{Probability distribution of the ratio of errors $r (\theta)$ between gapped and optimal cases for the sources in three astrophysical models. Blue and green contour indicate the single LISA detector and the joint Taiji-LISA configuration respectively.}
    \label{fig:tjls_catalog}
\end{figure*}

Tables \ref{tab:comparison} and \ref{table:detection_rate} quantitatively present the impact of data gaps on PE errors and detection rates. In Table \ref{tab:comparison}, the average $r(\theta)$ is used to represent the effects of data gaps. Since modes (2,1) and (4,4) are more challenging to be detected, and more comprehensive probability density distribution results are presented in Figures \ref{fig:tjtq_catalog} and \ref{fig:tjls_catalog}, so only the results for modes (2,2) and (3,3) are shown in Table \ref{tab:comparison} to avoid redundancy. On average, in the worst-case scenario, data gaps increase the PE errors of modes (2,2) and (3,3) to approximately 2.1 and 1.6 times their original values, respectively. Due to the similar arm lengths of Taiji and LISA, the effects of data gaps on these two detectors are comparable, with Taiji being slightly more affected. After joint observation, the influence of data gaps is significantly reduced, with $r(\theta)$ for mode (2,2) decreasing to around 1.5 and for mode (3,3) to around 1.3. The adverse impact of data gaps on PE accuracy during joint observation is approximately halved compared to that of a single detector. Since Taiji and LISA have lower sensitive frequency bands compared to TianQin, the Taiji-LISA joint observation configuration is more effective in reducing $r(\theta)$ than the Taiji-TianQin configuration. Finally, we observe that the real and imaginary parts of the same QNM have essentially identical $r(\theta)$, as they correspond to the same set of amplitudes.

Table \ref{table:detection_rate} shows the impact of data gaps on detection rates. Each value in the table represents the proportion of originally detectable sources that become undetectable due to data gaps. A larger value indicates a greater impact of data gaps. Joint observation significantly mitigates the negative effects of data gaps on detection rates.

In Figures \ref{fig:tjtq_catalog} and \ref{fig:tjls_catalog}, we calculate the relative change in PE errors. The horizontal axis in each panel represents $r(\theta)$, while the vertical axis shows the probability density distribution of $r(\theta)$, illustrating the likelihood of different levels of impact caused by data gaps. It can be observed that, due to the similar detector parameters of Taiji and LISA, the corresponding probability density distributions exhibit no significant differences. After joint observation, the probability density distribution in each panel shifts notably to the left, indicating a clear reduction in the effects of data gaps. Overall, although modes (3,3) and (4,4) have smaller average $r(\theta)$, their probability density distributions across different (2,2) and (2,1) models show a peak between 1.50 and 1.75, whereas modes (2,2) and (2,1) have a peak between 1.00 and 1.25.

\section{SUMMARY AND {OUTLOOK}}
\label{sec:conclusion}

Because the duration of ringdown signals is similar to the typical scale of data gaps, they may be significantly impacted by these gaps. This paper analyzes the effects of unscheduled gaps resulting in a 75$\%$ duty cycle. Using a method consistent with previous studies, we applied FIM analysis to assess how these gaps affect the accuracy of ringdown PE under worst-case conditions. We examined the characteristics of data gaps and their average impact using specific astrophysical models. Our results indicate that ringdown systems with larger remnant masses experience greater SNR degradation due to data gaps, while the impact varies minimally with the symmetric mass ratio. Based on calculations using astrophysical models, we find that, on average, the presence of data gaps increases the PE errors for modes (2,2) and (3,3) to approximately twice their values without gaps, while those for modes (3,3) and (4,4) increase to roughly 1.6 times. While Bayesian inference can provide more accurate results, its high computational demands limit its application to specific sources of interest, making it impractical for generating comprehensive average results across a sufficient number of ringdown signals. We plan to supplement our analysis with Bayesian inference in future work.

We analyzed two joint observation configurations, Taiji-TianQin and Taiji-LISA, and found that joint observation effectively reduces the influence of data gaps. When the signal detected by one detector is minimally affected by data gaps, the overall PE accuracy from joint observation can still closely approach the case without any data gaps, even if the other detector is significantly impacted. For scenarios where both detectors are affected by random data gaps to varying degrees, the average $r(\theta)$ is reduced from approximately 2.1 to 1.5 for the (2,2) mode, and from 1.6 to 1.3 for the (3,3) mode. The impact of gaps on detection rates is also mitigated through joint observation. Current results of PE are based on FIM calculations, and joint observation is expected to provide even better mitigation from the data analysis perspective. Due to the lower characteristic frequency band of the ringdown signal, the Taiji-LISA configuration, with its longer arms, is more effective in mitigating the impact of data gaps. These findings underscore the crucial role of joint observation.

The data processing approach in this paper involves evaluating signals with data gaps using optimal GW templates, providing a preliminary understanding of the effects of data gaps and joint observations. However, this is not the most precise assessment. On one hand, the smoothing applied to data gaps further reduces the SNR. On the other hand, knowledge of the existence of data gaps allows scientists to apply specific data-processing methods tailored to these gaps, potentially reducing their actual impact. Addressing data gaps by filling in missing data with AI or using templates with gap parameters in matched filter analysis is an ongoing area of research for our team. Moreover, data gaps not only result in signal loss but can also introduce phase inconsistencies between signal segments before and after the gap due to adjustments made to correct failures. This leads to the introduction of new parameters in the estimation process, and the impact of these new parameters on estimation accuracy requires further analysis. This will be a focus of our upcoming work.

\section*{Acknowledgement}

The authors thank Kallol Dey and Junjie Zhao for helpful discussion. This research is supported in part by the National Key R\&D Program of China, grant number 2020YFC2201300, and the National Natural Science Foundation of China, grant numbers 12035016, 12375058 and 12361141825.

\counterwithin{equation}{subsection}
\renewcommand{\thesubsection}{b}

\newpage

\bibliography{data_gap_onRD}

\end{document}